\documentclass[aps,pra,twocolumn,superscriptaddress,showpacs,amsmath,amssymb]{revtex4}

\usepackage{graphicx}
\usepackage{bm}
\usepackage{amssymb}

\begin{document}

\title{Characterizing optical chirality}

\author{Konstantin Y. Bliokh}
\affiliation{Applied Optics Group, School of Physics, National University of Ireland, Galway, Galway, Ireland}
\affiliation{Advanced Science Institute, RIKEN, Wako-shi, Saitama 351-0198, Japan}

\author{Franco Nori}
\affiliation{Advanced Science Institute, RIKEN, Wako-shi, Saitama 351-0198, Japan}
\affiliation{Physics Department, University of Michigan, Ann Arbor, Michigan 48109-1040, USA}


\begin{abstract}
We examine the recently introduced measure of chirality of a monochromatic optical field [Y. Tang and A. E. Cohen, Phys. Rev. Lett. \textbf{104}, 163901 (2010)] using the momentum (plane-wave) representation and helicity basis. Our analysis clarifies the physical meaning of the measure of chirality and unveils its close relation to the polarization helicity, spin angular momentum, energy density, and Poynting energy flow. We derive the operators of the optical chirality and of the corresponding chiral momentum, which acquire remarkably simple forms in the helicity representation.
\end{abstract}


\maketitle

\section{Introduction}

Optical chirality plays an important role in the optical sensing of biomolecules (see, e.g., [1]) and the interaction of light with chiral nanostructures or metamaterials [2]. The nonreciprocity in transmission or circular dichroism measures the chiral effects in the interaction of the optical field with the specimen. However, it is of both fundamental and practical interest to introduce a measure of the chirality of the optical field itself. Recently, Tang and Cohen [3] used a proposal by Lipkin [4] and introduced the local measure of the chirality of a nonparaxial monochromatic field. This will be called here optical \textit{chirality density} in order to emphasize its local nature. Together with the \textit{chirality flow density}, these quantities satisfy the continuity equation, akin to the Poynting theorem. A recent experiment [5] confirmed that this measure of optical chirality is meaningful.

One of the points of [4] was that the chirality density is not simply related to the local ellipticity (i.e., the degree of circular polarization of light in real space) but represents a more sophisticated characteristic which can take arbitrary large values when divided by the electric field energy density. Here we show that the chirality density is directly related to the \textit{helicity} of light, i.e., the degree of circular polarization in the \textit{momentum} (plane-wave) representation. We derive local and integral values of the chirality and chirality flow of a nonparaxial free field and show that the operators of energy and momentum of light, multiplied by the helicity, correspond to these quantities. This unveils the actual relation of the optical chirality to the polarization and sheds light on the similarity with the Poynting theorem.

\section{Basic equations}

To begin with, we consider monochromatic light in free space, characterized by the real electric and magnetic fields: $\bm{\mathcal E}({\bf r},t)$, $\bm{\mathcal H}({\bf r},t)$, and their standard complex representations: ${\bf E}\left( {\bf r} \right)$, ${\bf H}\left( {\bf r} \right)$, so that $\bm{\mathcal E}({\bf r},t) = {\mathop{\rm Re}\nolimits} \left[ {{\bf E}\left( {\bf r} \right)e^{ - i\omega t} } \right]$, ${\bm {\mathcal H}}({\bf r},t) = {\mathop{\rm Re}\nolimits} \left[ {{\bf H}\left( {\bf r} \right)e^{ - i\omega t} } \right]$. Using Gaussian units, we write the energy density $w$ and Poynting energy flow ${\bf s}$ [6]:
\begin{equation}\label{eqn:1}
w = {g \over 2}\left( {\bm{\mathcal E}^2  + \bm{\mathcal H}^2 } \right)~,
\end{equation}
\begin{equation}\label{eqn:2}
{\bf s} = c\,g \left( \bm{\mathcal E} \times \bm{\mathcal H}\right)~,
\end{equation}
where $g=(4\pi)^{-1}$. The chirality density $\chi$ and the corresponding chirality flow $\bm{\varphi}$ introduced in [3,4] read:
\begin{equation}\label{eqn:3}
\chi  = {c \over \omega }\,{g \over 2}\left[ {\bm{\mathcal E} \cdot \nabla  \times \bm{\mathcal E} + \bm{\mathcal H} \cdot \nabla  \times \bm{\mathcal H}} \right]~,
\end{equation}
\begin{equation}\label{eqn:4}
{\bm \varphi } = {{c^2 } \over \omega }\,{g \over 2}\left[ {\bm{\mathcal E} \times \left( {\nabla  \times \bm{\mathcal H}} \right) - \bm{\mathcal H} \times \left( {\nabla  \times \bm{\mathcal E}} \right)} \right]~.
\end{equation}
Compared to [3], we multiplied Eqs.~(3) and (4) by a constant $c/\omega$, in order to have the same dimensionality as the energy density and flow. The energy and chirality satisfy the continuity equations [3,6]:
\begin{equation}\label{eqn:5}
{{\partial w} \over {\partial t}} + \nabla  \cdot {\bf s} = 0~,
\end{equation}
\begin{equation}\label{eqn:6}
{{\partial \chi} \over {\partial t}} + \nabla  \cdot {\bm \varphi} = 0~.
\end{equation}

It is worth noticing that $\chi$ and ${\bm \varphi}$ are time-independent [4] (so that ${{\partial \chi} / {\partial t}} =0$), whereas $w$ and ${\bf s}$ possess oscillating terms [6]. Performing time averaging, we express these quantities via complex fields:
\begin{equation}\label{eqn:7}
\bar w = {g \over 4}\,{\mathop{\rm Re}\nolimits} \left( {{\bf E}^*  \cdot {\bf E} + {\bf H}^*  \cdot {\bf H}} \right)~,
\end{equation}
\begin{equation}\label{eqn:8}
{\bf \bar s} = c\,{g \over 2}\,{\mathop{\rm Re}\nolimits} \left( {{\bf E}^*  \times {\bf H}} \right)~,
\end{equation}
\begin{equation}\label{eqn:9}
\bar \chi  = \chi  =  - {g \over 2}\,{\mathop{\rm Im}\nolimits} \left( {{\bf E}^*  \cdot {\bf H}} \right)~,
\end{equation}
\begin{equation}\label{eqn:10}
{\bar {\bm \varphi} } = {\bm \varphi } = c\,{g \over 4}\,{\mathop{\rm Im}\nolimits} \left( {{\bf E}^*  \times {\bf E} + {\bf H}^*  \times {\bf H}} \right)~.
\end{equation}
Here we used Maxwell equations and wrote Eqs.~(7)--(10) in a form exhibiting a notable symmetry between the energy and chirality. While $w$ and ${\bf s}$ are respectively a scalar and a vector, $\chi$ and ${\bm \varphi}$ are a pseudoscalar and a pseudovectors changing their signs upon mirror reflections. After performing the time-averaging, the continuity equations (5) and (6) reduce to $\nabla  \cdot {\bf \bar s} = 0$ and  $\nabla  \cdot {\bm \varphi} = 0$.

\section{Helicity representation}

It is known that the chirality of an electromagnetic field is typically associated with the degree of circular polarization. At the same time, this quantity, known as helicity (i.e., the spin state of a photon), is well defined only for plane waves, i.e., in the momentum representation [7]. This is due to the transverse nature of the electromagnetic waves, for which the polarization is orthogonal to its wave vector. Therefore, one can naturally characterize the polarization of each plane wave in the spectrum, but it is difficult to characterize the local spatial polarization of a nonparaxial superposition of multiple plane waves -- in the generic case, the polarization has all three components and exhibits rather complicated features [8]. Recently, analyzing the energy flows in optical fields, we have shown that the dynamical characteristics (such as energy, momentum, and angular momentum) of a generic propagating electromagnetic field acquire a particularly clear and simple form in the helicity momentum representation [9]. Below we apply this representation to Eqs.~(7)--(10).

The Fourier plane-wave spectrum of the complex fields can be written as
\begin{equation}
\label{eqn:11}
\left\{ {\bf E}\left( {\bf r} \right), {\bf H}\left( {\bf r} \right) \right\} = {\alpha \over {2\pi }}\int {\left\{ {\bf \tilde E}\left( {\bf k} \right), {\bf \tilde H}\left( {\bf k} \right) \right\} e^{i{\bf k} \cdot {\bf r}} } d^2 {\bf k}~,
\end{equation}
where the integration is performed over the $(k_x, k_y)$ plane (for simplicity we assume propagating field and neglect evanescent waves) and the normalization factor $\alpha=\sqrt{2\omega /g}$ is introduced for convenience below. Now each complex Fourier amplitude can be represented as a sum of two circularly-polarized plane waves with well-defined helicities:
\begin{equation}\label{eqn:12}
{\bf \tilde E} = {\bf \tilde E}^+ + {\bf \tilde E}^- ~,~~~{\bf \tilde H} = {\bf \tilde H}^+ + {\bf \tilde H}^-~,
\end{equation}
where the helicities $\sigma=\pm 1$ correspond to the right-hand and left-hand circularly-polarized waves. The basic properties of the circular polarizations yield [7]
\begin{equation}\label{eqn:13}
{\bf \tilde H}^\sigma   =  - i\sigma {\bf \tilde E}^\sigma~,~~~{\bf \tilde E}^{\sigma *}  \times {\bf \tilde E}^\sigma   = i\sigma {{\bf k} \over k}\left| {{\bf \tilde E}^\sigma  } \right|^2~.
\end{equation}

Substituting the representation (11)--(13) and performing some calculations similar to those in [9,10] for the energy flows, we arrive at
\begin{equation}\label{eqn:14}
\bar w = {\omega  \over {\left( {2\pi } \right)^2 }}\sum\limits_{\sigma  =  \pm 1} {{\mathop{\rm Re}\nolimits} \int {d^2 {\bf k'}\int {d^2 {\bf k}\,e^{i {\bm \kappa} \cdot {\bf r}} \left( {{\bf \tilde E}^{\sigma * \prime} \cdot {\bf \tilde E}^\sigma  } \right)} } }~,
\end{equation}
\begin{equation}\label{eqn:15}
{\bf \bar s} = {{c\,\omega } \over {\left( {2\pi } \right)^2 }}\sum\limits_{\sigma  =  \pm 1} {{\mathop{\rm Im}\nolimits} \int {d^2 {\bf k'}\int {d^2 {\bf k}\,e^{i{\bm \kappa} \cdot {\bf r}} \sigma \left( {{\bf \tilde E}^{\sigma * \prime}  \times {\bf \tilde E}^\sigma  } \right)} } }~,
\end{equation}
\begin{equation}\label{eqn:16}
\chi  = {\omega  \over {\left( {2\pi } \right)^2 }}\sum\limits_{\sigma  =  \pm 1} {{\mathop{\rm Re}\nolimits} \int {d^2 {\bf k'}\int {d^2 {\bf k}\,e^{i{\bm \kappa} \cdot {\bf r}} \sigma \left( {{\bf \tilde E}^{\sigma * \prime} \cdot {\bf \tilde E}^\sigma  } \right)} } }~,
\end{equation}
\begin{equation}\label{eqn:17}
{\bm \varphi } = {{c\,\omega } \over {\left( {2\pi } \right)^2 }}\sum\limits_{\sigma  =  \pm 1} {{\mathop{\rm Im}\nolimits} \int {d^2 {\bf k'}\int {d^2 {\bf k}\,e^{i{\bm \kappa} \cdot {\bf r}} \left( {{\bf \tilde E}^{\sigma * \prime} \times {\bf \tilde E}^\sigma  } \right)} } }~,
\end{equation}
where ${\bm \kappa} = {{\bf k} - {\bf k'}}$ and ${\bf \tilde E}^{\sigma * \prime} \equiv {\bf \tilde E}^{\sigma *} \left( {{\bf k'}} \right)$. It is clear that in Eqs.~(14)--(17) the integrands of the energy and chirality quantities differ \textit{only} by the helicity $\sigma$, which flips upon mirror reflections and spatial inversion. Note also that Eqs.~(14)--(17) do \textit{not} contain interference cross-terms mixing different helicities -- this is a remarkable feature of the helicity representation diagonalizing quadratic field forms of Maxwell equations [9,10]. 

From Eqs.~(14) and (16) it immediately follows that the ratio of the chirality density $\chi$ to the energy density ${\bar w}$ cannot exceed 1 in absolute value:
\begin{equation}\label{eqn:18}
{\chi  \over {\bar w}} \, = \, {{\bar w^+   - \bar w^{\,-}} \over {\bar w^+ + \bar w^{\,-}}} \, \in \, \left[ { - 1,1} \right]~.
\end{equation}
Here $\bar w^\sigma$ is the energy density of the $\sigma$-polarized part of the field and the maximal chirality $\chi/{\bar w}=\sigma$ is achieved for a field composed of plane waves with the same helicity $\sigma$. It should be emphasized that in this case the actual local polarization of the optical field in real space (which results from interference of multiple plane waves propagating in different directions) can be rather complicated and far from being circular [8,10]. 

Our results clarify the examples given in the Supporting Online Materials to [3], where a superposition of two counter-propagating plane waves with different polarization was considered. One can note that the chirality density obtained for such field represents the sum of energy-weighted helicities of the partial waves. The enormously high response in the so-called superchiral fields was achieved in [3,5], because there the chirality efficiency was determined by the ratio of the chirality density to the \textit{electric} field energy density, which is only a part of the total energy density of the field. (Indeed, in the example [3] of counterpropagating orthogonally-polarized waves, the maxima of the electric energy density correspond to the minima of the magnetic energy density and vise versa.) 
Note also that in [5] the evanescent near-field modes were involved, which are beyond the scope of this work.

It is interesting to calculate the integral values of the energy, momentum, chirality, and corresponding `chiral momentum' of the field. The momentum density ${\bf p}$ differs from the energy flow ${\bf \bar s}$ by a factor of $c^2$ (${\bf p} = {\bf \bar s}/c^2$), and we introduce the corresponding \textit{chiral momentum} density: ${\bm \pi } = {\bm \varphi }/c^2$. For propagating beam-like fields the integral quantities per unit $z$-length are determined by the 2D spatial integration over the $(x,y)$ plane [9,11]:
\begin{eqnarray}\label{eqn:19}
\nonumber
W = \int {\bar w} \,d^2 {\bf r}~,~~~{\bf P} = \int {{\bf p}} \,d^2 {\bf r}~,\\
{\rm X} = \int \chi  \,d^2 {\bf r}~,~~~
{\bm \Pi } = \int {\bm \pi } \,d^2 {\bf r}~.
\end{eqnarray}
Performing this integration of Eqs.~(14)--(17) and using $\int {e^{i{\bm \kappa} \cdot {\bf r}} } d^2 {\bf r} = \left( {2\pi } \right)^2 \delta ^2 \left( {\bm \kappa} \right)$ along with Eq.~(13), we obtain
\begin{equation}\label{eqn:20}
W = \sum\limits_{\sigma  =  \pm 1} {\int {\omega \left| {{\bf \tilde E}^\sigma  } \right|^2 } }d^2 {\bf k}~,
\end{equation}
\begin{equation}\label{eqn:21}{\bf P} = \sum\limits_{\sigma  =  \pm 1} {\int {{\bf k} \left| {{\bf \tilde E}^\sigma  } \right|^2 } }d^2 {\bf k}~,
\end{equation}
\begin{equation}\label{eqn:22}
{\rm X} = \sum\limits_{\sigma  =  \pm 1} {\int {\sigma\omega \left| {{\bf \tilde E}^\sigma  } \right|^2 } }d^2 {\bf k}~,
\end{equation}
\begin{equation}\label{eqn:23}{\bm \Pi} = \sum\limits_{\sigma  =  \pm 1} {\int {\sigma{\bf k} \left| {{\bf \tilde E}^\sigma  } \right|^2 }} d^2 {\bf k}~.
\end{equation}
These equations allow a clear interpretation in terms of the quantum-like operators of the corresponding dynamical quantities [7,9,12]. Introducing the field state vector $\left| {{\bf \tilde E}^\sigma  } \right\rangle  = {\bf \tilde E}^\sigma$ and assuming the convolution defined as $\left\langle ~~ \right|\left. ~ \right\rangle  = \sum\limits_{\sigma  =  \pm 1} {\int {d^2 {\bf k}} }$, we see that Eqs.~(20)--(23) can be regarded as the expectation values ${\bf O} = \left\langle {{\bf \tilde E}^\sigma  } \right|{\bf \hat O}\left| {{\bf \tilde E}^\sigma  } \right\rangle$ of the following operators of energy, momentum, chirality, and chiral momentum:
\begin{eqnarray}\label{eqn:24}
\nonumber
\hat W = \omega~,~~~~{\bf \hat P} = {\bf k}~,\\
\hat {\rm X} = \sigma \omega~,~~~~{\bf \hat \Pi } = \sigma {\bf k}~.
\end{eqnarray}

Thus, using the operator formalism, the 4-pseudovector of the chirality and chiral momentum, $\left(\hat{\rm X},\hat{\bm \Pi}\right)$ represent just the usual energy-momentum $\left( {\omega ,{\bf k}} \right)$ multiplied by the pseudoscalar of helicity $\sigma$. This remarkably simple result unveils the actual physical meaning of the measure of chirality introduced in [3-5] and explains its connection with the polarization state and Poynting theorem. Importantly, the chiral momentum is closely related to the \textit{spin angular momentum} of light, which is represented by the operator ${\bf \hat \Sigma } = \sigma {\bf k}/k$ [9]. Hence, for monochromatic fields the chirality (22) and chiral momnetum (23) represent, up to constant factors, the averaged helicity of the field and its spin angular momentum:
\begin{equation}\label{eqn:25}
{\rm X} = \omega \langle \sigma \rangle~,~~~~{\bm \Pi} = \frac{\omega}{c} {\bf \Sigma }~.
\end{equation}
\[ \]

\section{Conclusion}

To summarize, we have examined the measure of the optical chirality for a free monochromatic optical field. Using the momentum representation (Fourier decomposition), we uncovered the close connections of the chirality density and chirality flow to the polarization helicity, energy density, and Poynting energy flow. The ratio of the chirality density to the energy density reaches its maximum absolute value for the fields composed of plane waves with the same well-defined helicity. At the same time, the local spatial polarization structure resulting from the interference of partial plane waves can be rather complicated. Finally, we have determined the integral values of the chirality and the corresponding chiral momentum and have revealed that the usual energy-momentum operator multiplied by the helicity underlies these quantities.

\textit{Note added.--} After submission of this work, a related paper [13] came to our attention. Similar quantities and conservation laws are considered there, but in the context of the Riemann-Silberstein vectors, which enables to deal with helicities in the coordinate representation [7b].

We acknowledge support from the European Commission (Marie Curie Action), Science Foundation Ireland (Grant No. 07/IN.1/I906),
LPS, NSA, ARO, DARPA, AFOSR, NSF grant No. 0726909, JSPSRFBR contract No. 09-02-92114, Grant-in-Aid for Scientific Research (S), MEXT Kakenhi on Quantum Cybernetics, and the Funding Program for Innovative R\&D on Science and Technology (FIRST).

\end{document}